\UseRawInputEncoding
\documentclass[%
 aps,
 prd,
 twocolumn,
 groupedaddress,  
 nofootinbib,
 amsmath,amssymb,
 floatfix,
]{revtex4-2}
\usepackage{natbib}
\usepackage{graphicx}
\usepackage{adjustbox}
\usepackage{dcolumn}
\usepackage{bm}
\usepackage[dvipsnames]{xcolor}
\usepackage{color, colortbl}
\usepackage{relsize}
\usepackage[mathlines]{lineno}
\usepackage{mathtools} 

\usepackage{stackengine}
\stackMath
\usepackage{xcolor}
\usepackage{dcolumn}
\usepackage{bm}
\usepackage{physics}
\usepackage{comment}
\usepackage{hyperref}
\usepackage{cleveref}
\usepackage{float}
\usepackage{soul}

\begin{document}

\title{Dark matter dynamical friction on binary stars in the faintest dwarf galaxies:\\ I.  Analytical approach}
\author{Nicolas Esser}
\email[]{nicolas.esser@ulb.be} 
\affiliation{Service de Physique Th\'eorique, Universit\'e libre de Bruxelles}
\affiliation{Brussels Laboratory of the Universe -- BLU-ULB \\Boulevard du Triomphe, CP225, 1050 Brussels, Belgium}
\date{\today}

\begin{abstract}
We study binary stars moving through a uniform dark matter background and experiencing dynamical friction. The centre-of-mass motion of the pairs is taken into account. We derive formulas and timescales for the secular evolution of the orbital parameters for both wide and close binaries. Applying these results to environments typical of the faintest ultra-faint dwarf galaxies, we show that some binaries undergo significant eccentricity oscillations while their semi-major axes gradually decrease over a Hubble time. At the population level, dynamical friction depletes wide binaries and increases the fraction of highly eccentric systems. Importantly, the effect \textit{is not} strongest at the highest dark matter densities, since the associated increase in velocity dispersion weakens friction. With future, more detailed theoretical studies and improving observational capabilities, binary stars may serve as a tool to probe the dark matter content of the faintest galaxies.

\end{abstract}

\maketitle

\section{Introduction}

The nature of dark matter (DM) remains one of the most pressing questions in astrophysics \cite{Strumia_2024}. Despite substantial observational evidence for DM on galactic and extragalactic scales, probing it on subgalactic scales remains challenging. Nevertheless, such investigations could provide significant insights into its fundamental properties \cite{Zavala_2019}.

Here, we explore the effect of dark matter on binary stellar systems, focusing on DM in the form of objects lighter than stars, including particle-like DM. Unlike massive compact DM objects, which can inject energy into binaries and possibly lead to their disruption \cite{Monroy_2014}, lighter DM can exert dynamical friction on stars, altering their orbital evolution and potentially producing other observable signatures. However, the effect of DM friction on binaries in the solar neighborhood was studied decades ago and found to be negligible \cite{Bekenstein_1990}.

But the discovery of tens of DM-dominated ultra-faint dwarf galaxies (UFDs) around the Milky Way over the past two decades has opened a new testbed for dark matter \cite{Simon_2019}. Thanks to ever-improving observational capabilities, progressively fainter systems continue to be discovered each year \cite{Mau_2020,Tan_2025,Cerny_2026}, and significant progress has recently been made in detecting and characterizing binaries within these galaxies \cite{Safarzadeh_2022,Shariat_2025,Muratore_2025,Qiu_2025}. This raises the question of the impact of DM dynamical friction on binary stars in such environments.

In this paper, we characterize the long-term secular evolution of binary orbital parameters driven by DM dynamical friction and apply our results to binary star populations in ultra-faint dwarf galaxies. We find that dynamical friction can significantly affect binary stars in the faintest UFDs over a Hubble time, shrinking wide binaries to tighter orbits and increasing the fraction of eccentric binaries. Interestingly, these effects are not strongest in UFDs with the highest DM densities, as the associated increase in velocity dispersion weakens dynamical friction. Binary stars in mildly DM-dominated environments are therefore more strongly affected than those in denser halos, where stars have higher velocities. Ultimately, binary stars themselves could serve as probes of dark matter in some of the faintest galaxies.

The rest of this paper is organized as follows.
After introducing the general framework for dynamical friction acting on binary stars in Sec.~\ref{sec:dyna_fric}, we analyze two opposite regimes: in Sec.~\ref{sec:wide}, we consider wide binaries, whose orbital velocities are small compared to their centre-of-mass (CoM) velocities; the opposite regime, corresponding to close binaries, is treated in Sec.~\ref{sec:close}. We derive analytical expressions and timescales for both regimes and apply them, in Sec.~\ref{sec:pop}, to a simple binary star population in an environment typical of the faintest UFDs. We conclude in Sec.~\ref{sec:conclusions}.

\section{Dynamical friction on binaries}
\label{sec:dyna_fric}
We use the standard Chandrasekhar formula to describe the dynamical friction exerted on a star of mass $m_i$ moving with velocity $\bm{v}_i$ through a DM background of uniform density $\rho_\text{dm}$. Assuming that the DM particles follow a Maxwellian velocity distribution with dispersion $\sigma$, the acceleration due to dynamical friction reads \cite{Chandra_1,Binney+Tremaine},
\begin{equation}
\bm{f}_i=-4\pi G^2\rho_\text{dm}m_i\ln\Lambda\ \mathcal{A}\left(\frac{v_i}{\sigma}\right)\frac{\bm{v}_i}{v_i^3},
\label{eq:dynafric}
\end{equation}
with $\mathcal{A}(X)$ a dimensionless monotonic function, described in Appendix \ref{app:functions}, such that $\mathcal{A}(X\rightarrow 0)\propto X^3$ and $\mathcal{A}(X\rightarrow\infty)=1$. One therefore recovers the standard results $f_i \propto v_i^{-2}$ for $v_i \gg \sigma$ and $f_i \propto v_i$ for $v_i \ll \sigma$. The Coulomb logarithm $\ln\Lambda$ has been assigned various expressions in the literature \cite{Binney+Tremaine,Kavanagh_2025}. We follow \cite{Binney+Tremaine} and adopt $\ln\Lambda=\ln\left[r_{1/2} v_i^2/(Gm_i)\right]=10$ for the remainder of this work. This value is appropriate for stars moving at a few km/s in ultra-faint dwarf galaxies with typical half-light radii $r_{1/2}\sim\mathcal{O}(10)\text{ pc}$. Because of the logarithm, even strong variations in the above parameters lead only to $\mathcal{O}(1)$ changes. The dynamical friction acceleration arises from the gravitational force generated by the wake of DM particles trailing a star as it moves through the background. The mass of this wake is proportional to the stellar mass; more massive stars generate larger wakes, which in turn slow them down more effectively. Consequently, the strength of the dynamical friction acceleration scales with $m_i$.

In a binary stellar system, the velocity of each component can be decomposed as $\bm{v}_i=\bm{v}_\text{cm}+\bm{v}_{\text{rel},i}$, where $\bm{v}_\text{cm}$ is the centre-of-mass velocity of the pair, and $\bm{v}_{\text{rel},i}$ is the relative velocity of star $i$ with respect to the CoM. The latter can be written as $\bm{v}_{\text{rel},1}=\dot{\bm{x}}m_2/M$ and $\bm{v}_{\text{rel},2}=-\dot{\bm{x}}m_1/M$, with $M=m_1+m_2$ the total mass of the system, and $\dot{\bm{x}}$ the relative velocity between the two stars, whose separation vector $\bm{x}$ points from the second star to the first. Using Eq.~\eqref{eq:dynafric}, the effect of dynamical friction on the global motion of the pair follows from the definition of the CoM velocity, $\bm{f}_\text{cm}=(m_1\bm{f}_1+m_2\bm{f}_2)/M$, while its effect on the internal orbital motion is $\bm{f}_\text{df}=\bm{f}_1-\bm{f}_2$, that is,
\begin{align}
\begin{split}
&\bm{f}_\text{df}=-4\pi G^2\rho_\text{dm}\ln\Lambda\\&\times\left[m_1\mathcal{A}\left(\frac{v_1}{\sigma}\right)\frac{\bm{v}_1}{v_1^3}-m_2\mathcal{A}\left(\frac{v_2}{\sigma}\right)\frac{\bm{v}_2}{v_2^3}\right].
\label{eq:orbital_evo}
\end{split}
\end{align}

The total internal acceleration of the system is given by $\ddot{\bm{x}}=\ddot{\bm{x}}_0 + \bm{f}_\text{df}$, where $\ddot{\bm{x}}_0=-GM \bm{x}/x^3$ is the standard two-body gravitational acceleration. The internal energy and angular momentum are defined as $E=\mu(\dot{x}^2/2-GM/x)$ and $\bm{L}=\mu\bm{x}\cross \dot{\bm{x}}$, from which one can derive their evolution under dynamical friction,
\begin{equation}
\dot{E} = \mu \dot{\bm{x}} \cdot \bm{f}_\text{df} \text{ and } \dot{\bm{L}} = \mu \bm{x} \times \bm{f}_\text{df},
\label{eq:EdotLdot}
\end{equation}
where $\mu = m_1 m_2 / M$ is the reduced mass. The dynamical friction force is very weak compared to the mutual gravitational attraction of the binary components. Its effect therefore operates on a timescale much longer than the orbital period. As a result, a single orbit is effectively Keplerian. One can then write $E=-GM\mu/(2a)$ and $L=\mu\sqrt{GMa(1-e^2)}$, with $a$ and $e$ the orbital semi-major axis and eccentricity of the reduced orbit. Because the evolution is slow, we are interested in the long-term, secular changes of the orbital parameters $E$ and $\bm{L}$. Since a single orbit remains Keplerian, the orbital average of any quantity $Q$ can be computed by integrating over the period, expressed in terms of the true anomaly $\theta$, via the relation (see e.g. \cite{Maggiore})
\begin{equation}
    \langle Q\rangle=\frac{(1-e^2)^{3/2}}{2\pi}\int_0^{2\pi}\frac{Q(\theta)}{\left(1+e\cos\theta\right)^2}d\theta.
\label{eq:orbitaverage}
\end{equation}
We will use the orbit-averaged evolution of the orbital parameters to characterize the long-term behavior of binary systems.

Two simple cases will be considered: wide and close binaries, defined here as systems whose internal orbital velocity is small or large, respectively, compared to their CoM velocity. The transition between these two regimes occurs for binaries with typical separation
\begin{equation}
    a\sim GM/v_\text{cm}^2\sim10^3\text{ AU}\left(\frac{M}{M_\odot}\right)\left(\frac{\text{km}/\text{s}}{v_\text{cm}}\right)^2.
    \label{eq:sep_wide_close}
\end{equation}

\section{Wide binaries}
\label{sec:wide}
The wider the binary, the smaller the orbital velocity. In the limit where $\dot x\ll v_\text{cm}$, expanding Eq.~\eqref{eq:orbital_evo} at first order leads to
\begin{align}
\begin{split}
    \bm{f}_\text{df}=&-\frac{4\pi G^2\rho_\text{dm}\ln\Lambda}{v_\text{cm}^3}\bigg[(m_1-m_2)\mathcal{A}\left(\frac{v_\text{cm}}{\sigma}\right)\bm{v}_\text{cm}  
    \\[5pt]&+2\mu\left(\mathcal{A}\left(\frac{v_\text{cm}}{\sigma}\right)\dot{\bm{x}}-\mathcal{B}\left(\frac{v_\text{cm}}{\sigma}\right)\frac{\dot{\bm{x}}\cdot\bm{v}_\text{cm}}{v_\text{cm}^2}\bm{v}_\text{cm}\right)\bigg],
    \label{eq:a_wide}
\end{split}
\end{align}
where $\mathcal{B}(X)$ is a dimensionless monotonic function, described in Appendix \ref{app:functions}, such that $\mathcal{B}(X\rightarrow0)\propto X^5$ and $\mathcal{B}(X\rightarrow\infty)=1$.
As indicated by the last term in Eq.~\eqref{eq:a_wide}, in the wide-binary limit the orbital evolution due to dynamical friction depends on the relative orientation between the CoM velocity $\bm{v}_\text{cm}$ and the orbital velocity $\dot{\bm{x}}$. We therefore need a fully general description of the Keplerian orbits, taking into account the orientation of the ellipse in three-dimensional space. This is usually done by writing $\bm{x}=r\,\bm{u}_r$ with 
\begin{equation}
\begin{array}{c}
r = \dfrac{a(1-e^2)}{1+e\cos\theta} \\[12pt]
\bm{u}_r =
\begin{bmatrix}
\cos\Omega\,\cos(\omega+\theta) - \cos i\,\sin\Omega\,\sin(\omega+\theta) \\[2pt]
\sin\Omega\,\cos(\omega+\theta) + \cos i\,\cos\Omega\,\sin(\omega+\theta) \\[2pt]
\sin i\,\sin(\omega+\theta)
\end{bmatrix},
\end{array}
\label{eq:orbital_elements}
\end{equation}
where $\theta$ is the true anomaly describing the time-dependent position of the stars, $i$ is the inclination of the orbital plane with respect to the $xy$-plane, $\Omega$ is the longitude of the ascending node, and $\omega$ is the argument of periapsis (see e.g. \cite{Poisson_Will,BMW}). The velocity $\dot{\bm{x}}$ can be computed from Eq.~\eqref{eq:orbital_elements} using $d/dt=\dot{\theta}\cdot d/d\theta$, with $\dot{\theta}=L/(\mu r^2)$.

With the help of Eqs.~\eqref{eq:EdotLdot} to \eqref{eq:orbital_elements}, one may evaluate $\langle \dot{E} \rangle$. Importantly, the zeroth-order part (i.e. the first line) of Eq.~\eqref{eq:a_wide} depends only on the CoM velocity, and not on the orbital position or velocity. Once incorporated into the expression for $\dot E$ and orbit-averaged, it therefore leads to a term proportional to $\langle \dot{\bm{x}} \rangle = 0$. The latter vanishes because the orbital velocity is large at small orbital radii, where the stars spend less time, and small at large radii, where they spend more time. When averaged over time, these contributions exactly cancel. The energy evolution is therefore dominated by the first-order corrections. Without loss of generality, we choose the CoM velocity to be oriented along the positive $z$-axis. One then finds
\begin{equation}
\hspace{-0.04cm}\langle \dot E\rangle=\frac{E}{\tau_{w,1}}\left[1-\frac{\mathcal{B}\left(v_\text{cm}/\sigma\right)}{\mathcal{A}\left(v_\text{cm}/\sigma\right)}\frac{\sin^2 i}{2}\left(1-\mathcal{F}(e)\cos2\omega\right)\right]
\label{eq:E_evo_wide}
\end{equation}
where we have defined the wide binary energy loss (i.e. first order) timescale as
\begin{align}
\begin{split}
&\tau_{w,1}=\left[\frac{16\pi G^2\rho_\text{dm}\mu \ln\Lambda}{v_\text{cm}^3}\mathcal{A}\left(\frac{v_\text{cm}}{\sigma}\right)\right]^{-1}\\[5pt]
&\sim4\times10^{9}\text{ yr}\left(\frac{10\text{ GeV}/\text{cm}^3}{\rho_\text{dm}}\right)\left(\frac{\max (v_\text{cm},\sigma)}{\text{km}/\text{s}}\right)^3.
\label{eq:tau_w1}
\end{split}
\end{align}
The numerical estimate assumes $m_1=m_2=M_\odot$, and the appearance of $\max(v_\mathrm{cm},\sigma)$ arises from considering the two limiting regimes of $\mathcal{A}(v_\text{cm}/\sigma)$. $\mathcal{F}(e)$ is a dimensionless monotonic function, described in Appendix \ref{app:functions}, such that $\mathcal{F}(0)=0$ and $\mathcal{F}(1)=1$. The bracketed factor in Eq.~\eqref{eq:E_evo_wide} can take values $\in[-2,1]$.

The evolution of the semi-major axis can be readily derived from its relation to the orbital energy, $\dot a = -a\dot E / E$, and can be either negative or positive. However, situations in which $a$ increases are rare, as they typically require a peculiar combination of a CoM velocity larger than the velocity dispersion, a large eccentricity, and specific values of $i$ and $\omega$, such that the bracketed factor in Eq.~\eqref{eq:E_evo_wide} becomes negative. For $i=0$ (a binary orbit perpendicular to the CoM velocity), one finds
\begin{equation}
a(t)=a_0\,\exp(-t/\tau_{w,1}),
\label{eq:analytical_evo_wide}
\end{equation}
where $a_0$ denotes the initial semi-major axis. Since $\tau_{w,1}$ is independent of $a_0$, the ratio $a(t)/a_0$ is also independent of $a_0$, thus implying a scale-invariant behaviour.

To fully characterize the evolution of the binary orbit, one must determine $\dot{\bm{L}}$ in addition to $\dot E$. This, however, is still insufficient; an ellipse has five degrees of freedom: two specifying the orbital plane, one its overall size, one its eccentricity, and one the orientation of its major and minor axes within the plane. Energy and angular momentum account for only four of these. The remaining degree of freedom, associated with the orientation of the axes, is therefore undetermined. This ambiguity is commonly resolved by introducing the eccentricity (or Laplace-Runge-Lenz) vector. Alternatively, the orbit can be parametrized by the set $\{a,e,i,\Omega,\omega\}$, which is often more convenient. In particular, the evolution of these elements under an external perturbation has already been derived in the literature (see e.g. \cite{Poisson_Will,BMW}). While an orbit average can always be performed, it is particularly simple when the perturbation is constant -- as is the case here at zeroth order in Eq.~\eqref{eq:a_wide} -- yielding
\begin{align}
\begin{split}&
\begin{dcases}
    \langle\dot e\rangle=-\frac{1}{\tau_{w,0}}\sqrt{\frac{a}{a_0}}\sqrt{1-e^2}\cos\omega\sin i\\[2pt]
    \langle\dot i\rangle=\frac{1}{\tau_{w,0}}\sqrt{\frac{a}{a_0}}\frac{e}{\sqrt{1-e^2}}\cos\omega\cos i\\[2pt]
    \langle\dot \Omega\rangle=\frac{1}{\tau_{w,0}}\sqrt{\frac{a}{a_0}}\frac{e}{\sqrt{1-e^2}}\frac{\sin\omega}{\tan i}\\[2pt]
    \langle\dot\omega\rangle=\frac{1}{\tau_{w,0}}\sqrt{\frac{a}{a_0}}\frac{\left(1-2e^2-\cos 2i\right)}{2\sqrt{1-e^2}}\frac{\sin\omega}{\sin i}.
    \label{eq:params_evo_wide}
\end{dcases}
\end{split}
\end{align}
As expected, $\langle \dot a \rangle = 0$ is also recovered at zeroth order. We again assumed that the CoM velocity is oriented along the positive $z$-axis.  We have defined the zeroth-order wide binary evolution timescale as
\begin{align}
\begin{split}
\hspace{-0.29cm}
\tau_{w,0}&=\left[\frac{6\pi G^{3/2}\rho_\text{dm}(m_1-m_2)\ln\Lambda\sqrt{a_0}}{v_\text{cm}^2\sqrt{M}}\mathcal{A}\left(\frac{v_\text{cm}}{\sigma}\right)\right]^{-1}\\[10pt]
&\sim4\times10^{9}\text{ yr}\left(\frac{10\text{ GeV}/\text{cm}^3}{\rho_\text{dm}}\right)\left(\frac{10^4\text{ AU}}{a_0}\right)^{1/2}\\[5pt]
&\hspace{1.7cm}\times\left(\frac{\text{km}/\text{s}}{v_\text{cm}}\right)\left(\frac{\max (v_\text{cm},\sigma)}{\text{km}/\text{s}}\right)^3.
\label{eq:tau_w0}
\end{split}
\end{align}
The numerical estimate assumes $m_1=1.5M_\odot$ and $m_2=0.5M_\odot$. Importantly, $\tau_{w,0}$ is proportional to the mass difference between the two stars, so the evolution of $\{e,i,\Omega,\omega\}$ is slower when this difference is small. In general, $\tau_{w,0} \propto \tau_{w,1}\sqrt{GM/a}/v_\mathrm{cm}$, indicating that, compared to the other orbital parameters, the evolution of $a$ is suppressed by one power of the orbital-to-CoM velocity ratio. However, for the adopted values, the numerical estimates in Eqs.~\eqref{eq:tau_w1} and \eqref{eq:tau_w0} are comparable, such that all orbital elements evolve on the same timescale.

\begin{figure}[t!]
\includegraphics[width=1.\columnwidth]{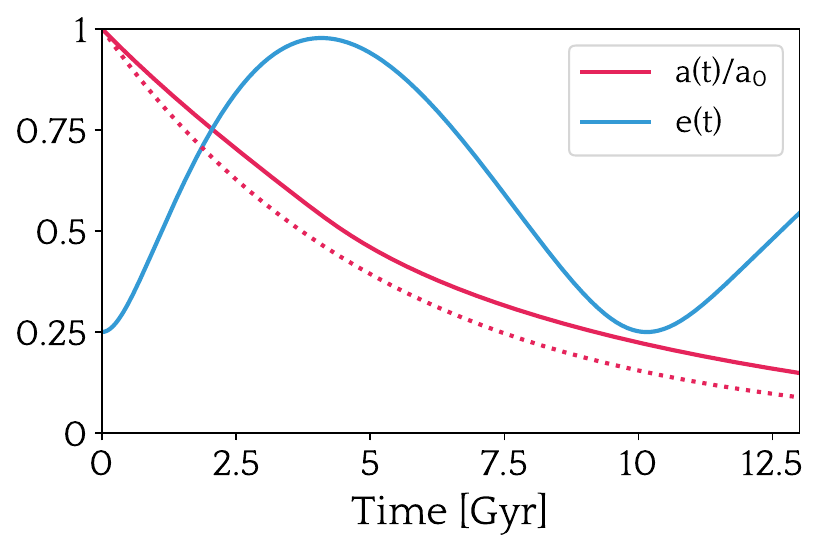}
\caption{\label{fig:evo_wide} Evolution of the semi-major axis (red) and eccentricity (blue) of a wide binary with $m_1 = 1.5 M_\odot$ and $m_2 = 0.5 M_\odot$, initial values $a_0 = 10^4$ AU and $e_0 = 0.25$, and CoM velocity $v_\text{cm} = 1$ km/s, in an environment with $\rho_\text{dm} = 10$ GeV/cm$^3$ and $\sigma = 1$ km/s. Solid curves show the full evolution obtained by integrating Eqs.~\eqref{eq:E_evo_wide} and \eqref{eq:params_evo_wide}, while the dotted curve shows the analytical solution from Eq.~\eqref{eq:analytical_evo_wide}. The orbital parameters $i$, $\Omega$, and $\omega$ also evolve but are not shown; changing their initial values leads to different evolutionary tracks.}
\end{figure}

Figure~\ref{fig:evo_wide} shows an example of the evolution of $a$ and $e$ for a wide binary system with $v_\text{cm}=1\text{ km}/\text{s}$ in a DM-dominated, low-velocity-dispersion environment with $\rho_\text{dm} = 10\text{ GeV}/\text{cm}^3$ and $\sigma = 1\text{ km}/\text{s}$. Solid curves show the results of numerically solving Eqs.~\eqref{eq:E_evo_wide} and \eqref{eq:params_evo_wide}, while the dotted curve corresponds to the idealised analytical solution from Eq.~\eqref{eq:analytical_evo_wide}. 
As seen in the figure, binaries can undergo large eccentricity oscillations, while their semi-major axes decrease by a factor of a few over a Hubble time under the adopted environment properties. The eccentricity evolution arises from differential drag: both stars move through the same DM background at nearly the CoM velocity, but since the dynamical friction acceleration depends on mass, the heavier star experiences a stronger drag, generating a torque on the pair that drives these oscillations. At high CoM velocities, the dynamical friction effect weakens because the stars move away more rapidly from their DM wake, diminishing the resulting gravitational drag.

The reduction of the semi-major axis of wide binaries in dwarf galaxies has already been studied in Refs.~\cite{Hernandez_2008,hernandez_2025}. However, these works considered that the radius of the sphere of influence of each binary -- corresponding to the typical size of the wakes generated in the DM background -- was equal to the distance between the binary components, an assumption that is not generally valid. Focusing on circular orbits, they found an evolution timescale for $a$ parametrically smaller than ours by a factor $\propto \sqrt{GM/a}/\sigma$, albeit with a larger numerical coefficient\footnote{Note two typos in Ref.~\cite{Hernandez_2008}. \textit{(i)} The value given for their scaling factor is incorrect; the correct value is $\alpha = 1.03$ \cite{hernandez_2025}; and \textit{(ii)} square roots are missing on the binary radii $R_0$ in the expression for the time immediately below their Eq.~(9).}. Overall, we find comparable values for the evolution timescale of $a$ for very wide binaries with separations $\sim 10^4 \text{ AU}$. Meanwhile, our scaling agrees, up to a correction in the Coulomb logarithm, with the results of Refs.~\cite{Gould_1991, Quinlan_1996} (see also \cite{Pani_2015}), which studied the effect of friction on binaries from first principles, without invoking the Chandrasekhar formula. 

\section{Close binaries}
\label{sec:close}
The Chandrasekhar formula encounters several problems for close binaries. The typical size $l$ of the DM wake generated by each star can be estimated by comparing the DM kinetic energy with the stellar gravitational potential, yielding $l \sim G m_i / \sigma^2$ \cite{Bekenstein_1990}. For close binaries, the orbital velocity satisfies $\dot{x} \sim \sqrt{GM/a} \gg v_{\rm cm} \sim \sigma$, implying $a \lesssim l$. In this regime, the wake generated by each star cannot be treated as affecting that star independently without also influencing its companion. Moreover, the energy extracted from the stars is transferred to the dark matter, which is heated and eventually ejected, leading over time to a depletion of the local DM background. While this effect is negligible for wide binaries, for which both the CoM velocity and the velocity dispersion greatly exceed the orbital velocity, it may become important in the opposite regime of close binaries, where the stars can complete many orbits without significant motion through the galaxy and without sufficient time for dark matter to replenish the local environment.

While aware of these limitations, we will nonetheless employ the Chandrasekhar formula for close binaries to obtain an estimate of the magnitude of its effects on such systems. The results presented below should therefore be regarded as idealized, and a more accurate treatment will ultimately be required. Given the substantial effort needed to model the feedback of friction on the DM halo -- even in simplified scenarios such as large mass-ratio binaries with negligible CoM motion \cite{Kavanagh_2025} -- we defer such a study to future work.

Close binaries have large orbital velocities. In the limit where $\dot x\gg v_\text{cm}$, the leading order of Eq.~\eqref{eq:orbital_evo} is
\begin{align}
\begin{split}
\bm{f}_\text{df}=&-\frac{4\pi G^2M^3\mu\rho_\text{dm}\ln\Lambda}{\dot x^3}\\[5pt]&\times\left[\frac{1}{m_1^3}\mathcal{A}\left(\frac{m_1\dot x}{M\sigma}\right)+\frac{1}{m_2^3}\mathcal{A}\left(\frac{m_2\dot x}{M\sigma}\right)\right]\dot{\bm{x}}.
\label{eq:a_tight}
\end{split}
\end{align}
Since Eq.~\eqref{eq:a_tight} depends only on the internal orbital velocity and not on the CoM velocity, the friction force is confined (at zeroth order) to the orbital plane, which therefore remains fixed in time. We can thus set $\Omega = i = 0$, i.e. take the orbital plane to coincide with the $xy$-plane. The direction of the angular momentum remains constant, and only its magnitude evolves in time. Given that the centre-of-mass velocity of a given binary is typically comparable to the velocity dispersion $\sigma$, one can further assume $\dot x\gg\sigma$, such that $\mathcal{A}(m_i\dot x/(M\sigma))\simeq1$. In this approximation, we can derive the evolution of $E$ and $L$ from Eqs.~\eqref{eq:EdotLdot}, \eqref{eq:orbitaverage} and \eqref{eq:a_tight} as

\begin{align}
\begin{split}&
\begin{dcases}
    \langle\dot E\rangle=\frac{E}{\tau_c}\left(\frac{a}{a_0}\right)^{3/2}\mathcal{K}(e,1)\\[5pt]
    \langle \dot L\rangle=-\frac{L}{2\tau_c}\left(\frac{a}{a_0}\right)^{3/2}\mathcal{K}(e,3),
\end{dcases}
\label{eq:E_and_L_evo_tight}
\end{split}
\end{align}
with $\mathcal{K}(e,n)$ a dimensionless monotonic function, described in Appendix \ref{app:functions}, such that $K(0,n)=1$, $K(e\rightarrow1,n)\rightarrow\infty$, and $\mathcal{K}(e,3)>\mathcal{K}(e,1)$. We have defined the close binary evolution timescale as
\begin{align}
\begin{split}
\tau_c&=\left[8\pi\rho_\text{dm}\ln\Lambda \sqrt{G}M^{3/2}\mu\left(\frac{1}{m_1^3}+\frac{1}{m_2^3}\right)a_0^{3/2}\right]^{-1}\\[10pt]
&\sim7\times10^{9}\text{ yr}\left(\frac{10\text{ GeV}/\text{cm}^3}{\rho_\text{dm}}\right)\left(\frac{100\text{ AU}}{a_0}\right)^{3/2},
\label{eq:tau_c}
\end{split}
\end{align}
where the numerical estimate assumes $m_1=m_2=M_\odot$.

The evolution of the semi-major axis and the eccentricity can be straightforwardly derived from Eq.~\eqref{eq:E_and_L_evo_tight} using the relations $\dot a=-a\dot E/E$ and $\dot e=-(1-e^2)/e\times[\dot E/(2E)+\dot L/L]$. One finds in particular $\langle \dot a\rangle=-a^{5/2}\mathcal{K}(e,1)/(\tau_c a_0^{3/2})$. In the case of circular orbits, i.e. $e=0$, the solution for the semi-major axis as a function of time is 
\begin{equation}
    a(t)=a_0\left(1+3t/2\tau_c\right)^{-2/3}.
    \label{eq:analytical_evo_tight}
\end{equation}

\begin{figure}[t!]
\includegraphics[width=1.\columnwidth]{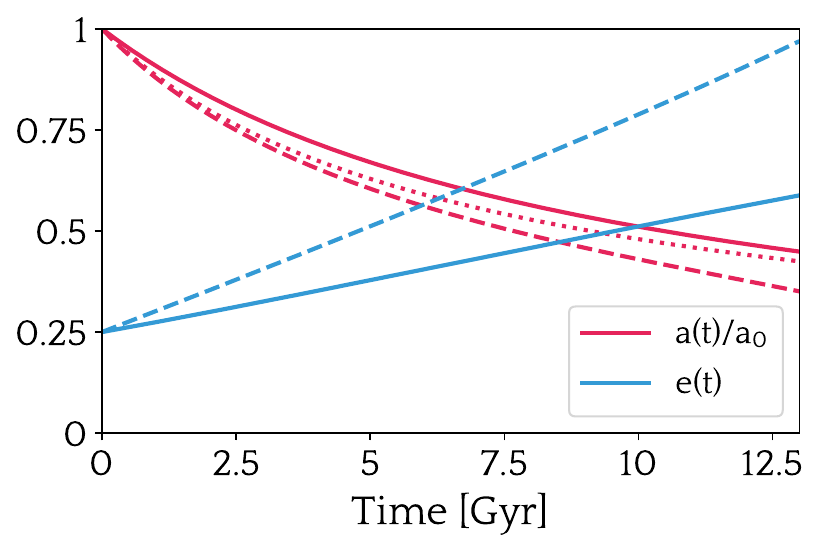}
\caption{\label{fig:evo_tight} Evolution of the semi-major axis (red) and eccentricity (blue) of a close binary of solar-mass stars with initial values $a_0=100$ AU and $e_0=0.25$, in a environment with $\rho_\text{dm}=10$ GeV$/$cm$^3$ and $\sigma=1$ km$/$s. Solid curves show the numerical integration of Eq.~\eqref{eq:a_tight}, dashed curves the solutions of Eq.~\eqref{eq:E_and_L_evo_tight}, an the dotted curve the analytical solution \eqref{eq:analytical_evo_tight}.}
\end{figure}

Figure \ref{fig:evo_tight} shows an example of the evolution of $a$ and $e$ for a close binary system in a DM-dominated, low-velocity-dispersion environment with $\rho_\text{dm} = 10$ GeV/cm$^3$ and $\sigma = 1$ km/s. The solid curves are obtained by directly integrating Eq.~\eqref{eq:a_tight}, which only assume $\dot x\gg v_\text{cm}$. The dashed curves are the numerical solutions of Eq.~\eqref{eq:E_and_L_evo_tight}, which further assume $\dot{x} \gg \sigma$. The dotted curve shows the analytical solution given by Eq.~\eqref{eq:analytical_evo_tight} for circular orbits.

Unlike in the wide-binary case, the evolution of the semi-major axis is not scale-invariant. For the environment properties considered here, a binary with an initial semi-major axis of $a_0=100\text{ AU}$ shrinks by a factor of $\sim 2$ over a Hubble time, whereas a pair with smaller initial semi-major axis would undergo negligible evolution, as indicated by Eq.~\eqref{eq:tau_c}. We also observe an increase in eccentricity. This arises because the friction force, which scales as $1/\dot{x}^2$, is stronger at apastron than at periastron \cite{Szolgyen_2022, ONeill_2024}. For the case shown in Fig.~\ref{fig:evo_tight}, Eq.~\eqref{eq:E_and_L_evo_tight} overestimates the increase in $e$ and the decrease in $a$, as the condition $\dot x\gg\sigma$ is only marginally satisfied. We verified that it agrees with Eq.~\eqref{eq:a_tight} when this condition is fully satisfied.

Note that larger eccentricities lead to a stronger decrease in $a$, as seen in Eq.~\eqref{eq:E_and_L_evo_tight}. In particular, $\mathcal{K}(e,1) \simeq 2$ for $e = 0.9$ and $\simeq 5$ for $e = 0.999$. This trend arises because the low orbital velocity near apastron in highly eccentric binaries enhances the friction. However, at very high eccentricities, the apastron velocity may become so small that the assumptions $\dot{x} \gg v_\text{cm}$ and $\dot{x} \gg \sigma$ break down, and the equations derived in this section are no longer applicable.

\section{Effect on a stellar population}
\label{sec:pop}
We consider, as a toy model, a population of binary stars with masses drawn from a Chabrier initial mass function \cite{Chabrier_2003}, initial eccentricities following the thermal distribution \cite{Heggie_1975}, and initial semi-major axes distributed according to a log-normal distribution \cite{Raghavan_2010}. We note that these distributions are inferred from observations of binary stars in the solar neighborhood, but may differ for stellar populations in UFDs.

Centre-of-mass velocities are drawn from a Maxwellian distribution with dispersion $\sigma$, and the relative orientation of the inner orbit with respect to the CoM velocity is picked at random. We only consider stars in the mass range $[0.1, 0.8]\,M_\odot$, as stars $\gtrsim 0.8\,M_\odot$ have reached the main-sequence turnoff in old stellar populations, while stars $\lesssim0.1\,M_\odot$ are difficult to resolve individually in UFD galaxies at $\mathcal{O}(10-100)$\,kpc distances, even with ultra-deep James Webb Space Telescope photometry \cite{Shariat_2025}.

We take as example parameters those inferred from the recently discovered faint stellar system DELVE 1 \cite{Mau_2020,Simon_2024}. Its stellar mass, estimated from its $V$-band magnitude of $M_V=-0.2$, is $M_\ast\simeq144\,M_\odot$. Conservative estimates of its line-of-sight velocity dispersion yield an upper bound of $\sigma_\text{los} = 1.2~\text{km/s}$, while its projected half-light radius is $r_h = 6.2~\text{pc}$. These quantities are related to their three-dimensional counterparts through $\sigma=\sqrt{3}\sigma_\text{los}$ and $r_{1/2}=4r_h/3$. Using the mass estimator for dispersion-supported stellar systems from \cite{Wolf_2010}, we obtain an upper bound on the total dynamical mass within the half-light radius of DELVE 1 of $M_{1/2}=4G^{-1}\sigma_\text{los}^2r_h\simeq8300\,M_\odot$. The resulting dynamical-to-stellar mass ratio, $\Upsilon_\text{dyn}=2M_{1/2}/M_\ast\simeq115$, is comparable to those inferred for some other DM-dominated UFDs \cite{Simon_2011}. Combining this mass estimate with the half-light radius yields a mean DM density of $\rho_\text{dm}\simeq130~\text{GeV}/\text{cm}^3$ within the central region.

\begin{figure*}[t]
\includegraphics[width=2.\columnwidth]{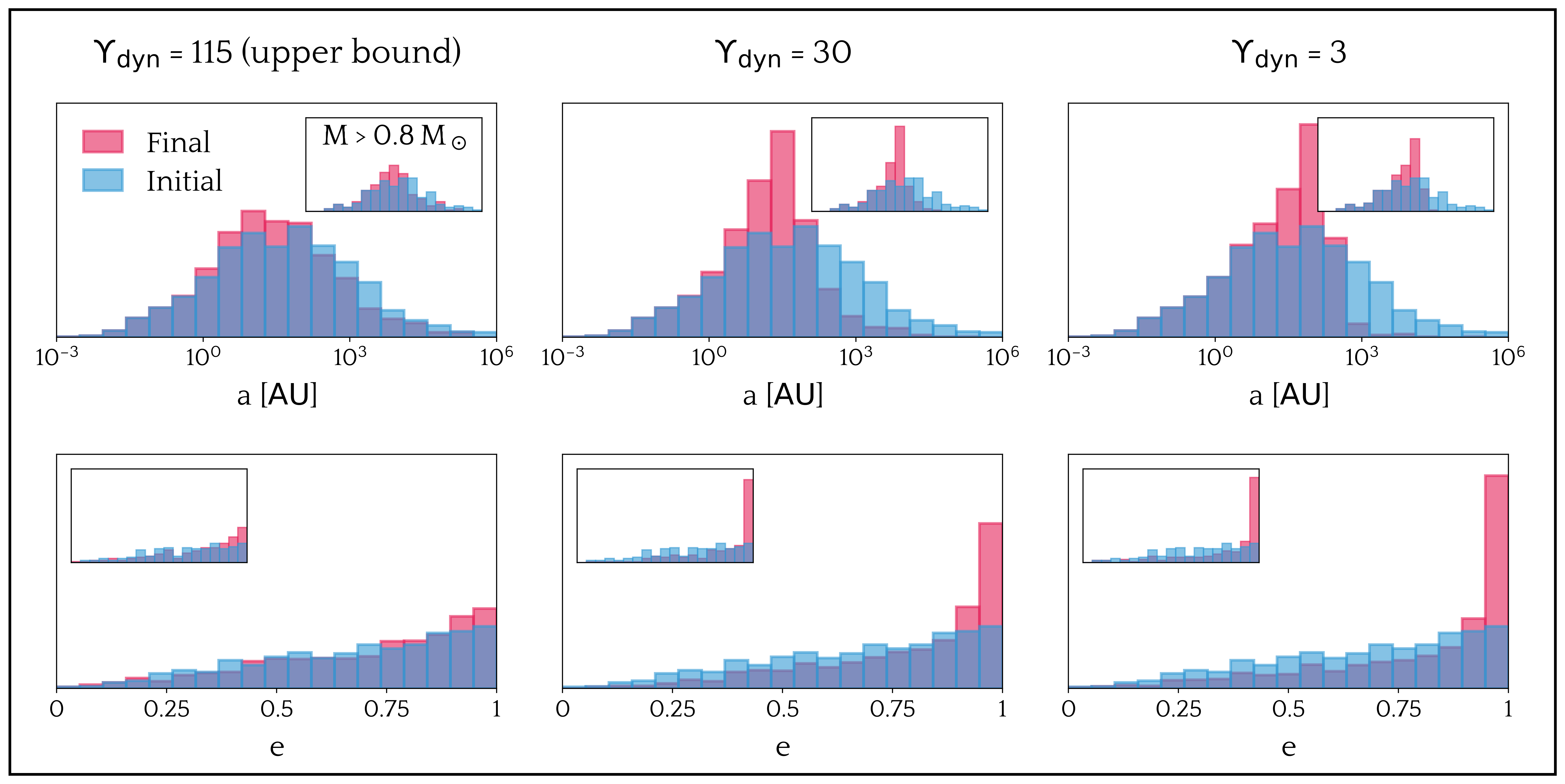}
\caption{Distribution of the initial (blue) and final (red) semi-major axes (upper panels) and eccentricities (lower panels) for a population of 1000 low-mass binaries evolved for 13 Gyr under dark matter dynamical friction, assuming a DELVE 1-like system with $\Upsilon_\text{dyn}=115$ (left), $\Upsilon_\text{dyn}=30$ (center) and $\Upsilon_\text{dyn}=3$ (right). The inset histograms highlight the 174 binaries with total mass $M > 0.8\,M_\odot$.}
\label{fig:pop_evo}
\end{figure*}

However, the actual velocity dispersion of DELVE 1 may be lower than the upper bound adopted above, implying a smaller dynamical mass and a lower degree of DM domination. To explore a broader range of possibilities for DELVE 1, we also consider two scenarios with lower dynamical-to-stellar mass ratios: an intermediate case with $\Upsilon_\text{dyn}=30$ and a mildly DM-dominated case with $\Upsilon_\text{dyn}=3$. These correspond to mean DM densities of $\rho_\text{dm}\simeq33$ and $2.3~\text{GeV}/\text{cm}^3$, and average velocity dispersions of $\sigma\simeq0.8$ and $0.3~\text{km/s}$, respectively \cite{Errani_2026}.

In all three cases, we numerically solve the general equation of binary evolution (Eq.~\eqref{eq:orbital_evo}) for $1000$ pairs\footnote{Although DELVE 1 does not contain this many stars, we adopt this sample size to ensure the statistical robustness of our results.}, in an orbit-averaged manner (Eq.~\eqref{eq:orbitaverage}). We verified that, in the appropriate limits, the results are equivalent to those obtained from the simpler evolution equations for wide and close binaries (Eqs.~\eqref{eq:E_evo_wide}, \eqref{eq:params_evo_wide}, and \eqref{eq:E_and_L_evo_tight}). Motivated by the old and relatively simple stellar populations of UFDs \cite{Simon_2019}, we evolve all stars for 13 Gyr.

In Fig.~\ref{fig:pop_evo}, we show the distribution in initial (blue) and final (red) semi-major axis (upper panels) and eccentricity (lower panels) of the population, for the three possible values of $\Upsilon_\text{dyn}$. DM friction effects are typically stronger when only considering the heavier stars in the population, as shown in the inset plots, which include only the 174 binaries in the sample with total mass $M > 0.8\,M_\odot$. We observe in all cases an overall increase in the number of high-eccentricity pairs and a rise in the number of binaries with semi-major axes around $\sim10-100$ AU, accompanied by a corresponding decrease in the number of low-eccentricity and wide binaries. 

While the overall decrease in semi-major axis across the population is expected from the energy dissipation associated with dynamical friction, the increase in the number of highly eccentric binaries is less intuitive and can be understood as follows. In wide binaries, as indicated by Eq.~\eqref{eq:E_evo_wide}, the semi-major axis $a$ typically decreases more rapidly as $e\to1$. Meanwhile, the evolution of $e$ slows as $a$ decreases, as follows from the first line of Eq.~\eqref{eq:params_evo_wide}. Consequently, eccentricity oscillations in wide binaries becomes effectively stalled at high eccentricity: as $e$ approaches unity, $a$ shrinks rapidly, suppressing further evolution of $e$ and keeping it close to unity. An increase in eccentricity is also expected for close binaries, as illustrated in Fig.~\ref{fig:evo_tight}.

More interestingly, we find that dynamical friction becomes more effective at lower $\Upsilon_\text{dyn}$: lower DM densities are associated with lower velocity dispersions (typically, $\sigma\propto\sqrt{\rho_\text{dm}}$), which in turn enhance the effect of friction, as illustrated by e.g. Eq.~\eqref{eq:tau_w1}. Thus, in systems with extremely low velocity dispersions, such as DELVE 1, a binary population strongly affected by dynamical friction does not necessarily imply a larger DM content.

Importantly, in the $\Upsilon_\text{dyn}=3$ case, the number of stars with semi-major axis $a>1000\text{ AU}$ goes from 174 to 12, while the number of stars with eccentricities $e>0.9$ goes from 202 to 503 following friction. These results highlight the substantial changes that dynamical friction can induce in the binary population, potentially detectable in future observations of even a small number of binaries in ultra-faint dwarf galaxies.

Another interesting result is the increase in the number of binary systems with small periastron passages. For example, in the initial population, there are 263 pairs with periastron distances $< 1\text{ AU}$. After evolving the population in the $\Upsilon_\text{dyn}=3$ case, and accounting for the combined effects of eccentricity oscillations in wide binaries and eccentricity growth in close binaries, we find that over the age of the galaxy a total of 476 pairs experience passages below 1 AU. Approximately $20\%$ of the binaries in the sample therefore reach sub-AU separations at some point because of dark matter friction. This may lead to tidal interactions, potentially inducing dynamical and chemical effects within the pair members and increasing the fraction of blue stragglers in UFDs \cite{Mathieu_2025}. However, further theoretical work is needed before firm conclusions can be drawn.

\section{Conclusion}
\label{sec:conclusions}

We have studied in detail the effect of dark matter dynamical friction on binary stars, accounting for the center-of-mass motion of the binaries within the dark matter halo. We derived analytical expressions and timescales for the evolution of the orbital parameters, and applied these results to a simple population of binaries.

We found that, in environments typical of ultra-faint dwarf galaxies, wide binaries may undergo strong eccentricity oscillations on $\sim$ Gyr timescales, while their semi-major axes gradually shrinks over a Hubble time. For close binaries, eccentricity tends to increase and semi-major axis decrease, although these effects depend on the initial separation and become negligible for sufficiently tight systems. This evolution typically leads to a depletion of wide binaries and a larger fraction of high eccentricity systems. It also results in more stars undergoing small periastron passages. A potential consequence is an increased number of binaries experiencing tidal interactions. More detailed theoretical studies will be needed to quantify this effect accurately. 

Several improvements to our analysis remain to be addressed. First, the applicability of the Chandrasekhar formula to close binaries must be clarified, as the DM wakes generated by the individual stars cannot be treated independently in this regime. A key objective is therefore to characterize dynamical friction in such systems using dedicated simulations that account for overlapping wakes and feedback on the DM background \cite{Kavanagh_2025}. Additional physical effects should also be incorporated to fully describe the dynamical evolution of binaries in UFDs, including galactic tides, three-body encounters, and migration toward the galactic centre driven by DM friction. In Appendix~\ref{app:timescales}, we estimate the associated timescales and find that galactic tidal forces may dominate for binaries with separations $\gtrsim \text{a few} \times 10^3\ \text{AU}$, highlighting the need for a more detailed investigation. Furthermore, the presence of specific DM and stellar density profile \cite{splawska_2026} and of a non-smooth DM component, such as massive subhalos \cite{Penarrubia_2024,Penarrubia_2025}, may influence our results and also warrant further study. Finally, adopting a more realistic initial distribution for the stellar population would provide a more complete and robust assessment. In particular, our simplified model considers only pairs of main-sequence stars, whereas stars in UFDs may also have compact-object companions, whose mass function is not accounted for in our analysis. Addressing all these issues will most likely require dedicated N-body simulations, which we leave to future work.

While we applied our calculations to ultra-faint dwarf galaxies, other heavily DM-dominated systems include DM spikes that may form around supermassive black holes \cite{Gondolo_1999,Bertone_2024}. In such spikes, the DM density can reach $\gtrsim 10^6\ \text{GeV}/\text{cm}^3$, although the velocity dispersion is higher than in UFDs. DM friction will influence binaries in these environments, but the tidal effects of the black hole cannot be neglected, requiring a coupled analysis of both phenomena. Nevertheless, these systems could produce interesting observable signatures in the binary population.

DELVE 1 was chosen as a representative system for our analysis; however, new faint systems are discovered every year. Their classification often requires multiple observations combining chemical and dynamical information. One such recently discovered object is the faint star cluster UNIONS 1 \cite{Smith_2024}. Although initially thought to be the most DM-dense galaxy ever identified \cite{Errani_2024}, subsequent studies suggest that it likely does not contain a significant amount of DM \cite{Cerny_2025} (but see also \cite{Adams_2026}). Our results indicate that dynamical friction is strongest in small systems with moderate DM dominance and extremely low velocity dispersion. As in the case of DELVE 1, if UNIONS 1 falls into this category, it could contain virtually no wide binaries, while most close binaries would have high eccentricities. The identification and characterization of even a few binaries in such a system could therefore provide valuable constraints on its DM content.

Upcoming surveys and instruments will enable the discovery of tens of new faint systems around the Milky Way (see e.g. \cite{Delve,Mutlu_Pakdil_2021,Manwadkar_2022,Tsiane_2025}) and allow more precise measurements of the binaries they contain (see e.g. \cite{Cooper_2023,skuladottir_2023,via_project}). Together, these developments underscore the need for further theoretical studies of the dynamical evolution of stars in the faintest Milky Way satellites, as well as dedicated observational searches for binaries, which may provide valuable insight into their nature and dark matter content.

\acknowledgements
Artificial intelligence tools were used solely to improve the language and structure of the manuscript. It is a pleasure to thank Peter Tinyakov for the many valuable discussions, Rapha\"el Errani for his insights and thoughtful exchanges and Marine Prunier for her helpful comments on the manuscript. The author is a FRIA grantee of the Fonds de la Recherche Scientifique-FNRS. Tag: ULB-TH/26-04.

\bibliographystyle{apsrev4-2} 

\bibliography{bibli}

@BOOK{Binney+Tremaine,
       author = {{Binney}, James and {Tremaine}, Scott},
        title = "{Galactic Dynamics: Second Edition}",
         year = 2008,
       adsurl = {https://ui.adsabs.harvard.edu/abs/2008gady.book.....B},
      publisher = {Princeton University Press}
}

@article{Hernandez_2008,
   title={The tightening of wide binaries in dSph galaxies through dynamical friction as a test of the dark matter hypothesis},
   volume={387},
   ISSN={1365-2966},
   url={http://dx.doi.org/10.1111/j.1365-2966.2008.13373.x},
   DOI={10.1111/j.1365-2966.2008.13373.x},
   number={4},
   journal={Monthly Notices of the Royal Astronomical Society},
   publisher={Oxford University Press (OUP)},
   author={Hernandez, X. and Lee, William H.},
   year={2008},
   month=jul, pages={1727–1734} }

@ARTICLE{hernandez_2025,
       author = {{Hernandez}, Xavier and {Kroupa}, Pavel},
        title = "{Dynamical Friction Constraints on the Dark Matter Hypothesis Across Astronomical Scales}",
      journal = {Universe},
         year = 2025,
        month = nov,
       volume = {11},
       number = {11},
          eid = {367},
        pages = {367},
          doi = {10.3390/universe11110367},
       adsurl = {https://ui.adsabs.harvard.edu/abs/2025Univ...11..367H}
}

@article{Kavanagh_2025,
  title = {Sharpening the dark matter signature in gravitational waveforms. II. Numerical simulations},
  author = {Kavanagh, Bradley J. and Karydas, Theophanes K. and Bertone, Gianfranco and Di Cintio, Pierfrancesco and Pasquato, Mario},
  journal = {Phys. Rev. D},
  volume = {111},
  issue = {6},
  pages = {063071},
  numpages = {17},
  year = {2025},
  month = {Mar},
  publisher = {American Physical Society},
  doi = {10.1103/PhysRevD.111.063071},
  url = {https://link.aps.org/doi/10.1103/PhysRevD.111.063071}
}

@article{Bertone_2024,
    author = "Bertone, Gianfranco and Wierda, A. Renske A. C. and Gaggero, Daniele and Kavanagh, Bradley J. and Volonteri, Marta and Yoshida, Naoki",
    title = "{Toward a realistic description of dark matter overdensities around black holes}",
    primaryClass = "astro-ph.CO",
    doi = "10.1103/5nnf-8fz9",
    journal = "Physical Review D",
    volume = "112",
    number = "4",
    pages = "043537",
    year = "2025"
}

@book{BMW,
  title        = {Fundamentals of Astrodynamics},
  author       = {Bate, Roger and Mueller, Donald and White, Jerry},
  year         = {1971},
  publisher    = {Dover Publications}
}

@BOOK{Maggiore,
       author = {{Maggiore}, Michele},
        title = "{Gravitational Waves: Volume 1: Theory and Experiments}",
         year = 2007,
         publisher = "Oxford University Press",
          doi = {10.1093/acprof:oso/9780198570745.001.0001},
       adsurl = {https://ui.adsabs.harvard.edu/abs/2007gwte.book.....M}
}

@ARTICLE{Bekenstein_1990,
       author = {{Bekenstein}, Jacob D. and {Zamir}, Ran},
        title = "{Dynamical Friction in Binary Systems}",
      journal = {The Astrophysical Journal},
         year = 1990,
        month = aug,
       volume = {359},
        pages = {427},
          doi = {10.1086/169075},
       adsurl = {https://ui.adsabs.harvard.edu/abs/1990ApJ...359..427B}
}

@ARTICLE{Pani_2015,
       author = {{Pani}, Paolo},
        title = "{Binary pulsars as dark-matter probes}",
      journal = {Physical Review D},
         year = 2015,
        month = dec,
       volume = {92},
       number = {12},
          eid = {123530},
        pages = {123530},
          doi = {10.1103/PhysRevD.92.123530},
 primaryClass = {astro-ph.HE},
       adsurl = {https://ui.adsabs.harvard.edu/abs/2015PhRvD..92l3530P}
}

@book{Poisson_Will,
  title        = {Gravity: Newtonian, Post-Newtonian, Relativistic},
  author       = {Poisson, Eric and Will, Clifford M.},
  year         = {2014},
  publisher    = {Cambridge University Press},
  doi          = {10.1017/CBO9781139507486},
  url          = {https://doi.org/10.1017/CBO9781139507486}
}

@ARTICLE{Gould_1991,
       author = {{Gould}, Andrew},
        title = "{Binaries in a Medium of Fast Low-Mass Objects}",
      journal = {The Astrophysical Journal},
         year = 1991,
        month = sep,
       volume = {379},
        pages = {280},
          doi = {10.1086/170502},
       adsurl = {https://ui.adsabs.harvard.edu/abs/1991ApJ...379..280G}
}

@article{Quinlan_1996,
   title={The dynamical evolution of massive black hole binaries I. Hardening in a fixed stellar background},
   volume={1},
   ISSN={1384-1076},
   url={http://dx.doi.org/10.1016/S1384-1076(96)00003-6},
   DOI={10.1016/s1384-1076(96)00003-6},
   number={1},
   journal={New Astronomy},
   publisher={Elsevier BV},
   author={Quinlan, Gerald D.},
   year={1996},
   month=jul, pages={35–56} }

@article{Szolgyen_2022,
   title={Eccentricity evolution in gaseous dynamical friction},
   volume={513},
   ISSN={1365-2966},
   url={http://dx.doi.org/10.1093/mnras/stac1294},
   DOI={10.1093/mnras/stac1294},
   number={4},
   journal={Monthly Notices of the Royal Astronomical Society},
   publisher={Oxford University Press (OUP)},
   author={Sz\"olgy\'en, Akos and MacLeod, Morgan and Loeb, Abraham},
   year={2022},
   month=may, pages={5465–5473} }

@article{ONeill_2024,
doi = {10.3847/1538-4357/ad7250},
url = {https://doi.org/10.3847/1538-4357/ad7250},
year = {2024},
month = {oct},
publisher = {The American Astronomical Society},
volume = {974},
number = {2},
pages = {216},
author = {O{\textquotesingle}Neill, David and D{\textquotesingle}Orazio, Daniel J. and Samsing, Johan and Pessah, Martin E.},
title = {Gaseous Dynamical Friction on Elliptical Keplerian Orbits},
journal = {The Astrophysical Journal}
}

@article{Esser_2025,
    author = "Esser, Nicolas",
    title = "{Primordial black hole-star binaries via dynamical friction}",
    primaryClass = "astro-ph.HE",
    doi = "10.1103/p98j-1jjz",
    journal = "Physical Review D",
    volume = "112",
    number = "4",
    pages = "043021",
    year = "2025"
}

@ARTICLE{Heggie_1975,
       author = {{Heggie}, D.~C.},
        title = "{Binary evolution in stellar dynamics.}",
      journal = {Monthly Notices of the Royal Astronomical Society},
         year = 1975,
        month = dec,
       volume = {173},
        pages = {729-787},
          doi = {10.1093/mnras/173.3.729},
       adsurl = {https://ui.adsabs.harvard.edu/abs/1975MNRAS.173..729H}
}

@article{Raghavan_2010,
   title={A SURVEY OF STELLAR FAMILIES: MULTIPLICITY OF SOLAR-TYPE STARS},
   volume={190},
   ISSN={1538-4365},
   url={http://dx.doi.org/10.1088/0067-0049/190/1/1},
   DOI={10.1088/0067-0049/190/1/1},
   number={1},
   journal={The Astrophysical Journal Supplement Series},
   publisher={American Astronomical Society},
   author={Raghavan, Deepak and others},
   year={2010},
   month=aug, pages={1–42} }

@ARTICLE{Chabrier_2003,
       author = {{Chabrier}, Gilles},
        title = "{Galactic Stellar and Substellar Initial Mass Function}",
      journal = {Publications of the Astronomical Society of the Pacific},
         year = 2003,
        month = jul,
       volume = {115},
       number = {809},
        pages = {763-795},
          doi = {10.1086/376392},
 primaryClass = {astro-ph},
       adsurl = {https://ui.adsabs.harvard.edu/abs/2003PASP..115..763C}
}

@article{Simon_2024,
    author = "Simon, Joshua D. and others",
    title = "{Eridanus III and DELVE 1: Carbon-rich Primordial Star Clusters or the Smallest Dwarf Galaxies?}",
    primaryClass = "astro-ph.GA",
    reportNumber = "FERMILAB-PUB-24-1019-PPD",
    doi = "10.3847/1538-4357/ad85dd",
    journal = "The Astrophysical Journal",
    volume = "976",
    pages = "256",
    year = "2024"
}

@ARTICLE{Mau_2020,
       author = {{Mau}, S. and others},
        title = "{Two Ultra-faint Milky Way Stellar Systems Discovered in Early Data from the DECam Local Volume Exploration Survey}",
      journal = {The Astrophysical Journal},
         year = 2020,
        month = feb,
       volume = {890},
       number = {2},
          eid = {136},
        pages = {136},
          doi = {10.3847/1538-4357/ab6c67},
 primaryClass = {astro-ph.GA},
       adsurl = {https://ui.adsabs.harvard.edu/abs/2020ApJ...890..136M}
}

@ARTICLE{Wolf_2010,
       author = {{Wolf}, Joe and others},
        title = "{Accurate masses for dispersion-supported galaxies}",
      journal = {Monthly Notices of the Royal Astronomical Society},
         year = 2010,
        month = aug,
       volume = {406},
       number = {2},
        pages = {1220-1237},
          doi = {10.1111/j.1365-2966.2010.16753.x},
 primaryClass = {astro-ph.CO},
       adsurl = {https://ui.adsabs.harvard.edu/abs/2010MNRAS.406.1220W}
}

@article{Shariat_2025,
doi = {10.1088/1538-3873/ae11d2},
url = {https://doi.org/10.1088/1538-3873/ae11d2},
year = {2025},
month = {oct},
publisher = {The Astronomical Society of the Pacific},
volume = {137},
number = {10},
pages = {104103},
author = {Shariat, Cheyanne and others},
title = {Wide Binaries in an Ultra-faint Dwarf Galaxy: Discovery, Population Modeling, and A Nail in the Coffin of Primordial black hole Dark Matter},
journal = {Publications of the Astronomical Society of the Pacific}
}

@article{Strumia_2024,
	title = {Dark matter},
	volume = {1},
	author = {Cirelli, Marco and Strumia, Alessandro and Zupan, Jure},
	journal = {SciPost Physics Reviews},
	year = {2026},
	publisher = {SciPost},
	url = {https://scipost.org/10.21468/SciPostPhysRev.1}
}

@Article{Zavala_2019,
AUTHOR = {Zavala, Jes\'us and Frenk, Carlos S.},
TITLE = {Dark Matter Haloes and Subhaloes},
JOURNAL = {Galaxies},
VOLUME = {7},
YEAR = {2019},
NUMBER = {4},
ARTICLE-NUMBER = {81},
URL = {https://www.mdpi.com/2075-4434/7/4/81},
PAGES = {81-135},
ISSN = {2075-4434},
DOI = {10.3390/galaxies7040081}
}

@article{Safarzadeh_2022,
   title={A Statistical Detection of Wide Binary Systems in the Ultrafaint Dwarf Galaxy Reticulum II},
   volume={930},
   ISSN={1538-4357},
   url={http://dx.doi.org/10.3847/1538-4357/ac626e},
   DOI={10.3847/1538-4357/ac626e},
   number={1},
   journal={The Astrophysical Journal},
   publisher={American Astronomical Society},
   author={Safarzadeh, Mohammadtaher and Simon, Joshua D. and Loeb, Abraham},
   year={2022},
   month=may, pages={54} }

@article{Muratore_2025,
   title={Exploring the ultra-faint dwarf Boötes I using JWST and HST: Metallicity distribution and binaries},
   volume={708},
   ISSN={1432-0746},
   url={http://dx.doi.org/10.1051/0004-6361/202558215},
   DOI={10.1051/0004-6361/202558215},
   journal={Astronomy \& Astrophysics},
   publisher={EDP Sciences},
   author={Muratore, F. and others},
   year={2026},
   month=Apr, pages={A100} }

@ARTICLE{Qiu_2025,
       author = {{Qiu}, Tian and others},
        title = "{The Binary Fraction of Stars in the Dwarf Galaxy Ursa Minor via Dark Energy Spectroscopic Instrument}",
      journal = {The Astrophysical Journal},
         year = 2026,
        month = mar,
       volume = {1000},
       number = {1},
          eid = {33},
        pages = {33},
          doi = {10.3847/1538-4357/ae422d},
       adsurl = {https://ui.adsabs.harvard.edu/abs/2026ApJ..1000...33Q}
}

@article{Monroy_2014,
   title={THE END OF THE MACHO ERA, REVISITED: NEW LIMITS ON MACHO MASSES FROM HALO WIDE BINARIES},
   volume={790},
   ISSN={1538-4357},
   url={http://dx.doi.org/10.1088/0004-637X/790/2/159},
   DOI={10.1088/0004-637x/790/2/159},
   number={2},
   journal={The Astrophysical Journal},
   publisher={American Astronomical Society},
   author={Monroy-Rodr\'iguez, Miguel A. and Allen, Christine},
   year={2014},
   month=jul, pages={159} }

@article{Simon_2019,
   title={The Faintest Dwarf Galaxies},
   volume={57},
   DOI={10.1146/annurev-astro-091918-104453},
   number={1},
   journal={Annual Review of Astronomy and Astrophysics},
   publisher={Annual Reviews},
   author={Simon, Joshua D.},
   year={2019},
   pages={375–415} }

@ARTICLE{Tan_2025,
       author = {{Tan}, C.~Y. and others},
        title = "{Ultra-faint Milky Way Satellites Discovered in Carina, Phoenix, and Telescopium with DELVE Data Release 3}",
      journal = {The Astrophysical Journal},
         year = 2026,
        month = mar,
       volume = {1000},
       number = {1},
          eid = {46},
        pages = {46},
          doi = {10.3847/1538-4357/ae43e2},
       adsurl = {https://ui.adsabs.harvard.edu/abs/2026ApJ..1000...46T}
}

@article{Penarrubia_2024,
    author = {Pe{\~n}arrubia, Jorge and Errani, Rapha{\"e}l and Walker, Matthew G. and Gieles, Mark and Boekholt, Tjarda C. N.},
    title = "{Capture of field stars by dark substructures}",
    primaryClass = "astro-ph.GA",
    doi = "10.1093/mnras/stae1961",
    journal = "Monthly Notices of the Royal Astronomical Society.",
    volume = "533",
    number = "3",
    pages = "3263--3282",
    year = "2024"
}

@article{Penarrubia_2025,
    author = {Pe{\~n}arrubia, Jorge and Errani, Rapha{\"e}l and Vitral, Eduardo and Walker, Matthew G.},
    title = "{Gravothermal expansion of dwarf spheroidal galaxies heated by dark subhaloes}",
    doi = "10.1093/mnras/staf1800",
    journal = "Monthly Notices of the Royal Astronomical Society",
    volume = "544",
    number = "2",
    pages = "2557--2576",
    year = "2025"
}

@article{Gondolo_1999,
    author = "Gondolo, Paolo and Silk, Joseph",
    title = "{Dark matter annihilation at the galactic center}",
    reportNumber = "MPI-PHT-99-10, OUAST-99-9",
    doi = "10.1103/PhysRevLett.83.1719",
    journal = "Physical Review Letters",
    volume = "83",
    pages = "1719--1722",
    year = "1999"
}

@ARTICLE{Chandra_1,
       author = {{Chandrasekhar}, S.},
        title = "{Dynamical Friction. I. General Considerations: the Coefficient of Dynamical Friction.}",
      journal = {The Astrophysical Journal},
         year = 1943,
        month = mar,
       volume = {97},
        pages = {255},
          doi = {10.1086/144517},
       adsurl = {https://ui.adsabs.harvard.edu/abs/1943ApJ....97..255C}
}

@ARTICLE{Smith_2024,
       author = {{Smith}, Simon E.~T. and others},
        title = "{The Discovery of the Faintest Known Milky Way Satellite Using UNIONS}",
      journal = {The Astrophysical Journal},
         year = 2024,
        month = jan,
       volume = {961},
       number = {1},
          eid = {92},
        pages = {92},
          doi = {10.3847/1538-4357/ad0d9f},
 primaryClass = {astro-ph.GA},
       adsurl = {https://ui.adsabs.harvard.edu/abs/2024ApJ...961...92S}
}

@article{Errani_2024,
   title={Ursa Major III/UNIONS 1: The Darkest Galaxy Ever Discovered?},
   volume={965},
   ISSN={1538-4357},
   url={http://dx.doi.org/10.3847/1538-4357/ad2267},
   DOI={10.3847/1538-4357/ad2267},
   number={1},
   journal={The Astrophysical Journal},
   publisher={American Astronomical Society},
   author={Errani, Raphaël and Navarro, Julio F. and Smith, Simon E. T. and McConnachie, Alan W.},
   year={2024},
   month=apr, pages={20} }

@ARTICLE{Cerny_2025,
       author = {{Cerny}, William and others},
        title = "{No Observational Evidence for Dark Matter Nor a Large Metallicity Spread in the Extreme Milky Way Satellite Ursa Major III/UNIONS 1}",
      journal = {The Astrophysical Journal Letters},
         year = 2026,
        month = mar,
       volume = {999},
       number = {1},
          eid = {L8},
        pages = {L8},
          doi = {10.3847/2041-8213/ae29b8},
       adsurl = {https://ui.adsabs.harvard.edu/abs/2026ApJ...999L...8C}
}

@ARTICLE{Delve,
       author = {{Drlica-Wagner}, A. and others},
        title = "{The DECam Local Volume Exploration Survey: Overview and First Data Release}",
      journal = {The Astrophysical Journal Supplement Series},
         year = 2021,
        month = sep,
       volume = {256},
       number = {1},
          eid = {2},
        pages = {2},
          doi = {10.3847/1538-4365/ac079d},
 primaryClass = {astro-ph.GA},
       adsurl = {https://ui.adsabs.harvard.edu/abs/2021ApJS..256....2D}
}

@article{Tsiane_2025,
   title={Predictions for the Detectability of Milky Way Satellite Galaxies and Outer-Halo Star Clusters with the Vera C. Rubin Observatory},
   volume={8},
   ISSN={2565-6120},
   url={http://dx.doi.org/10.33232/001c.142072},
   DOI={10.33232/001c.142072},
   journal={The Open Journal of Astrophysics},
   publisher={Maynooth University},
   author={Tsiane, Kabelo and others},
   year={2025},
   month=jul }

@article{Mutlu_Pakdil_2021,
   title={Resolved Dwarf Galaxy Searches within ∼5 Mpc with the Vera Rubin Observatory and Subaru Hyper Suprime-Cam*},
   volume={918},
   ISSN={1538-4357},
   url={http://dx.doi.org/10.3847/1538-4357/ac0db8},
   DOI={10.3847/1538-4357/ac0db8},
   number={2},
   journal={The Astrophysical Journal},
   publisher={American Astronomical Society},
   author={Mutlu-Pakdil, Burçin and others},
   year={2021},
   month=sep, pages={88} }

@article{Manwadkar_2022,
   title={Forward-modelling the luminosity, distance, and size distributions of the Milky Way satellites},
   volume={516},
   ISSN={1365-2966},
   url={http://dx.doi.org/10.1093/mnras/stac2452},
   DOI={10.1093/mnras/stac2452},
   number={3},
   journal={Monthly Notices of the Royal Astronomical Society},
   publisher={Oxford University Press (OUP)},
   author={Manwadkar, Viraj and Kravtsov, Andrey V},
   year={2022},
   month=sep, pages={3944–3971} }

@misc{via_project,
  title        = "{VIA Project}",
  howpublished = "\url{https://via-project.org/}"
}

@article{skuladottir_2023,
  author    = {Sk\'ulad\'ottir, \'A. and others},
  title     = {The 4MOST Survey of Dwarf Galaxies and their Stellar Streams (4DWARFS)},
  journal   = {The Messenger},
  volume    = {190},
  pages     = {19--21},
  month     = {March},
  year      = {2023},
  doi       = {10.18727/0722-6691/5304},
  url       = {https://doi.org/10.18727/0722-6691/5304}
}

@article{Cooper_2023,
   title={Overview of the DESI Milky Way Survey},
   volume={947},
   ISSN={1538-4357},
   url={http://dx.doi.org/10.3847/1538-4357/acb3c0},
   DOI={10.3847/1538-4357/acb3c0},
   number={1},
   journal={The Astrophysical Journal},
   publisher={American Astronomical Society},
   author={Cooper, Andrew P. and others},
   year={2023},
   month=apr, pages={37} }

@ARTICLE{Errani_2025,
       author = {{Errani}, Rapha{\"e}l and {Pe{\~n}arrubia}, Jorge and {Walker}, Matthew G.},
        title = "{Stellar Mass Segregation in Dark Matter Halos}",
      journal = {The Astrophysical Journal},
         year = 2025,
        month = nov,
       volume = {993},
       number = {2},
          eid = {160},
        pages = {160},
          doi = {10.3847/1538-4357/ae0328},
       adsurl = {https://ui.adsabs.harvard.edu/abs/2025ApJ...993..160E}
}

@article{splawska_2026,
    author = {Splawska, Sofia L. and Errani, Rapha{\"e}l and Pe{\~n}arrubia, Jorge and Walker, Matthew G.},
    title = "{Tailored Mass Estimators for Milky Way Dwarf Spheroidals}",
    doi = "10.3847/1538-4357/ae9a81",
    journal = "The Astrophysical Journal",
    volume = "1008",
    number = "2",
    pages = "191",
    year = "2026"
}

@article{Adams_2026,
  author  = {Adams, Tim R. and Brewer, Brendon J. and Lewis, Geraint F.},
  title   = {A Bayesian Exploration of The Mass of the Ursa Major III: Kinematics, Rotation and Their Influence on the Mass to Light Ratio},
  journal = {The Open Journal of Astrophysics},
  volume  = {9},
  month   = feb,
  year    = {2026},
  url     = {https://doi.org/10.33232/001c.158197}
}

@ARTICLE{Cerny_2026,
       author = {{Cerny}, William and others},
        title = "{Discovery of the Distant, Ultra-faint Milky Way Satellite Aquarius IV with the Vera C. Rubin Observatory Early Data Preview 2}",
      journal = {Research Notes of the American Astronomical Society},
         year = 2026,
        month = aug,
       volume = {10},
       number = {8},
          eid = {239},
        pages = {239},
          doi = {10.3847/2515-5172/ae993b},
       adsurl = {https://ui.adsabs.harvard.edu/abs/2026RNAAS..10..239C}
}

@ARTICLE{Simon_2011,
       author = {{Simon}, Joshua D. and others},
        title = "{A Complete Spectroscopic Survey of the Milky Way Satellite Segue 1: The Darkest Galaxy}",
      journal = {The Astrophysical Journal},
         year = 2011,
        month = may,
       volume = {733},
       number = {1},
          eid = {46},
        pages = {46},
          doi = {10.1088/0004-637X/733/1/46},
       adsurl = {https://ui.adsabs.harvard.edu/abs/2011ApJ...733...46S}
}

@article{Errani_2026,
    author = {Errani, Rapha{\"e}l and Esser, Nicolas and Pe{\~n}arrubia, Jorge and Walker, Matthew G.},
    title = "{Collisional Dynamics of Stars and Dark Matter in Ultrafaint Galaxies}",
    doi = "10.3847/1538-4357/ae819c",
    journal = "Astrophys. J.",
    volume = "1007",
    number = "1",
    pages = "42",
    year = "2026"
}

@ARTICLE{Mathieu_2025,
       author = {{Mathieu}, Robert D. and {Pols}, Onno R.},
        title = "{Blue Stragglers and Friends: Initial Evolutionary Pathways in Close Low-Mass Binaries}",
      journal = {Annual Review of Astronomy and Astrophysics},
         year = 2025,
        month = aug,
       volume = {63},
       number = {1},
        pages = {467-512},
          doi = {10.1146/annurev-astro-071221-054402},
       adsurl = {https://ui.adsabs.harvard.edu/abs/2025ARA&A..63..467M}
}

\appendix

\section{Dimensionless functions}
\label{app:functions}
Here, we define all the dimensionless monotonic functions appearing in the main text.

The first such functions introduced are
\begin{align}
\begin{split}
&\mathcal{A}\left(X\right)=\text{erf}\left(\frac{X}{\sqrt{2}}\right)-\sqrt{\frac{2}{\pi}}X\exp\left(-\frac{X^2}{2}\right)\\[5pt]
&\mathcal{B}(X)=3\, \mathcal{A}\left(X\right)-\sqrt{\frac{2}{\pi}}X^3\exp\left(-\frac{X^2}{2}\right).
\end{split}
\end{align}
Taking the lower and upper limits of these functions, one finds $\mathcal{A}(X\rightarrow0)=\sqrt{2/\pi}\ X^3/3$, $\mathcal{B}(X\rightarrow0)=\sqrt{2/\pi}\ X^5/5$ and $\mathcal{A}(X\rightarrow\infty)=\mathcal{B}(X\rightarrow\infty)=1$. Note that in the main text we use $X\equiv v_i/\sigma$, differing from Ref.~\cite{Binney+Tremaine}, where $X \equiv v_i/(\sqrt{2}\sigma)$ and $\mathcal{A}(X)$ is redefined accordingly.

We also use
\begin{equation}
\mathcal{F}(e)=\frac{2-e^2-2\sqrt{1-e^2}}{e^2},
\end{equation}
which is such that $F(0)=0$ and $F(1)=1$.

We finally introduce
\begin{equation}
\mathcal{K}(e,n)=\frac{(1-e^2)^{\frac{n+3}{2}}}{2\pi}\int\limits_0^{2\pi}\frac{(1+e\cos\theta)^{-2}}{(1+e^2+2e\cos\theta)^{n/2}}d\theta
\end{equation}
which is such that $\mathcal{K}(e,3)>\mathcal{K}(e,1)$. It also satisfies $\mathcal{K}(0,n)=1$ and diverges as $\mathcal{K}(e\rightarrow1,n)\rightarrow\infty$. In particular, carefully taking the limits of the integral yields in the zero-eccentricity limit
\begin{equation}
\lim_{e\rightarrow0}\mathcal{K}(e,n)\simeq1+\frac{1}{4}n(2+n)e^2,
\end{equation}
while in the high-eccentricity limit
\begin{align}
\begin{split}
     &\lim_{e\rightarrow1}\mathcal{K}(e,1)\simeq\frac{2}{\pi}\left[-1+\ln\left(\frac{8}{1-e}\right)\right]\\
     &\lim_{e\rightarrow1}\mathcal{K}(e,3)\simeq\frac{8}{\pi}\left(\frac{1}{1-e}\right).
\end{split}
\end{align}
The divergence of $\mathcal{K}(e,1)$ is logarithmic and therefore mild. Note that one must be cautious in the large-$e$ limit: Eq.~\eqref{eq:E_and_L_evo_tight}, which makes use of $\mathcal{K}(e,n)$, eventually becomes non-physical because the apastron velocity becomes very small at high eccentricities, reaching $\dot x|_\text{apastron} \lesssim v_\mathrm{cm}$, which contradicts the approximation used in the close binary case.

\section{Timescales comparison}
\label{app:timescales}
Here, we compare the timescales associated with the effect of DM dynamical friction on the internal motion of wide binaries -- e.g., $\tau_{w,1}$ (Eq.~\eqref{eq:tau_w1}) and $\tau_{w,0}$ (Eq.~\eqref{eq:tau_w0}) -- with those of other relevant mechanisms that may influence stellar dynamics in the typical UFD environments we consider, namely galactic tides, stellar encounters, and DM dynamical friction acting on the CoM of the pairs.

\subsection{Galactic tides}

Galactic tidal forces induce a precession of the elliptic plane of binary systems, which may either compete with or act in concert with the effect of DM friction. The tidal potential of a binary in a host galaxy of mass $\mathcal{M}$ and radius $\mathcal{R}$ is $U_\text{tide} \sim G \mathcal{M}a^2/ \mathcal{R}^{3}$. Comparing this with the two-body potential of the binary, $U_\text{pair} \sim G M / a$, gives a measure of the strength of the tidal perturbations. This strength is proportional to the ratio of the binary's internal orbital timescale to the tidal timescale, such that
\begin{equation}
    \tau_\text{tide}\sim\frac{M}{\mathcal{M}}\left(\frac{\mathcal{R}}{a}\right)^3\frac{2\pi a^{3/2}}{\sqrt{GM}}\sim10^{10}\text{ yr}\left(\frac{10^3\text{AU}}{a}\right)^{3/2}
\end{equation}
where the numerical estimate assumes $M=2\,M_\odot$ and adopts the half-light radius and, conservatively, the upper bound on the mass of DELVE 1.

\subsection{Stellar encounters}

Close encounters with passing stars can transfer momentum to binary systems, thereby altering their orbits. For wide binaries, stellar encounters can be treated using the impulse approximation \cite{Binney+Tremaine}, since the typical relative velocity between stars in UFDs, $\sigma_\text{rel}\simeq\sqrt{2}\sigma$, is much larger than the internal orbital velocity of such systems. The encounter rate of a binary with perturbers of impact parameter $b$ is $dN/dbdt = 2\pi b n_\ast \sigma_{\rm rel}$, where $n_\ast$ is the local stellar number density. In the impulse regime, a single fly-by of a perturbing star of mass $m_p$ induces a change in the relative velocity of the binary components of order $\Delta v \sim \pi G m_p / (b \sigma_{\rm rel})$, and therefore a (specific) energy kick $\Delta E \sim  \Delta v^2/2$. The cumulative effect of multiple encounters leads the the rate $\dot{E} = \int_{b_m}^{b_M} dN/dbdt \cdot \Delta E \,db$, where $b_m$ and $b_M$ are the minimum and maximum impact parameters for weak impulsive encounters. Computing the integral leads to $\dot{E}\sim\pi^3n_\ast G^2m_p^2\ln\lambda/\sigma_\text{rel}$, with $\lambda=b_M/b_m$ \cite{Esser_2025}. A comparison with the binding energy of the pair gives the timescale
\begin{equation}
     \tau_\text{s.e.}\sim \frac{8}{\pi^3\ln\lambda}\frac{\sigma}{GMan_\ast}\sim10^{11}\text{ yr}\left(\frac{10^3\text{AU}}{a}\right),
\end{equation}
where we set $m_p = M/2$. The numerical estimate assumes $\ln\lambda = 2$, $M = 2\,M_\odot$, and adopts the upper bound on the velocity dispersion inferred for DELVE 1, along with its stellar number density. The latter is estimated from the system's luminosity ($M_V = -0.2$), implying it contains roughly 250 stars within its half-light radius, which corresponds to a density of $n_\ast \simeq 0.1~\text{pc}^{-3}$.

\subsection{Friction on the CoM}
Another relevant timescale is the time taken by the pair to migrate towards the centre of the galaxy (or sink) due to dynamical friction acting on the centre-of-mass. Using $\dot E_\text{cm}=\bm{f}_\text{cm}\cdot\bm{v}_\text{cm}$ (see Sec.~\ref{sec:dyna_fric}), one can derive the CoM energy loss rate in the limit of wide binaries. Comparing this with the CoM energy of the pair, $E_\text{cm}\sim G\mathcal{M}M/\mathcal{R}$, yields the corresponding sinking timescale,
\begin{equation}
\tau_{\text{sink}}\sim\left[\frac{4\pi G\rho_\text{dm}\mathcal{R}\ln\Lambda}{v_\text{cm}}\frac{(m_1^2+m_2^2)}{\mathcal{M}M}\mathcal{A}\left(\frac{v_\text{cm}}{\sigma}\right)\right]^{-1}.
\end{equation}
For solar-mass stars with $v_\text{cm}=\sigma$ in an environment typical of DELVE 1 (using again the upper bound on its dynamical mass), one finds $\tau_\text{sink}\sim3\times10^9\text{ yr}$.

\subsection{Comparison}

\begin{figure}[t!]
\includegraphics[width=\columnwidth]{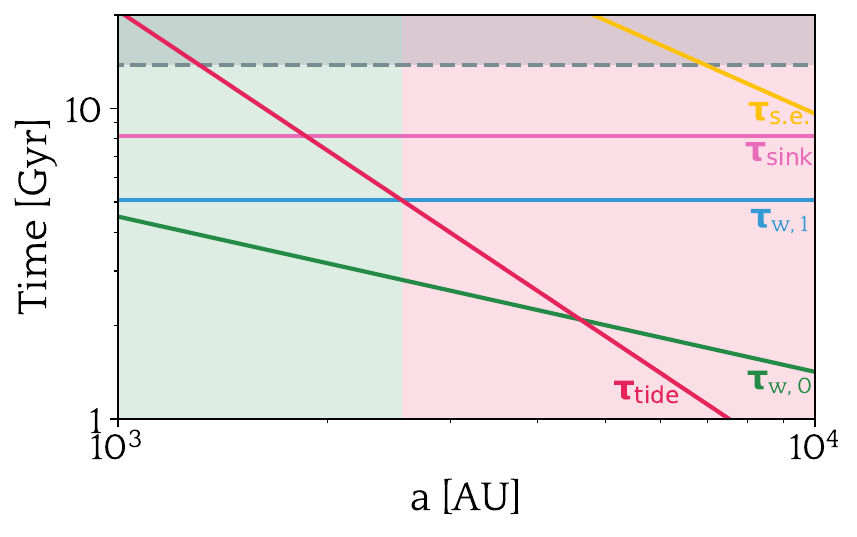}
\caption{\label{fig:timescales} 
Comparison of various dynamical timescales for wide binaries in an UFD galaxy. The upper grey region corresponds to times longer than the age of the Universe. The left green region indicates where DM dynamical friction dominates the long-term evolution of wide binaries. The right red region marks where at least one of the two wide-binary timescales exceeds the tidal timescale.}
\end{figure}

In Fig.~\ref{fig:timescales} we compare the various timescales described in the previous subsections with the wide binaries timescales introduced in Eqs.~\eqref{eq:tau_w1} and \eqref{eq:tau_w0}, as a function of the semi-major axis $a$ and adopting environmental parameters corresponding to the upper bound inferred for DELVE 1. The primary and secondary stellar masses are set to $0.8\,M_\odot$ and $0.4\,M_\odot$. As shown in the figure, galactic tides may dominate over internal DM friction in binaries wider than $\sim 2000\ \text{AU}$. They should therefore be carefully considered in any future, more detailed analysis of wide binary evolution in UFD galaxies. 

Stellar encounters and dynamical friction on the centre-of-mass have timescales longer than $\tau_{w,1}$ and $\tau_{w,0}$. Nevertheless, the sinking timescale of the binary pair as a whole remains shorter than the age of the Universe by a factor of a few, so stars are expected to migrate slightly toward the centres of UFDs. This mechanism could also potentially help constrain the DM content of faint galaxies, although it may be challenging due to the unknown formation sites of individual stars. Since the sinking time is also mass-dependent, it may lead to mass segregation and warrants further study (see also \cite{Errani_2025,Errani_2026}).

Finally, note that we neglected any eccentricity dependence in all timescales, but a more careful analysis should account for it as well.

\end{document}